\documentstyle[prl,aps,multicol,psfig]{revtex}
\begin{document}
\draft

\title{Mean-density Bogoliubov description of inhomogeneous 
Bose-condensed gases}

\author{John A. Vaccaro }
\address{
          Department of Physical Sciences,
          University of Hertfordshire,
          College Lane,
          Hatfield  AL10 9AB,
          United Kingdom.
}

\date{\today}
\maketitle

\begin{abstract}
A mean-density description of spatially-inhomogeneous Bose-condensed
gases based on Bogoliubov's method is introduced.  The description
assumes only a large mean atomic density and so remains valid when the
mean field collapses due to phase diffusion.  A spread in the number
of particles in the condensate is shown to lead to an anomalous
coupling between the condensate and excited modes.  This coupling is
due to the dependence of the condensate spatial wavefunction on
particle number and it could, in principle, be used for reducing
particle fluctuations in the condensate.
\end{abstract}

\pacs{PACS number(s): 03.75.Fi, 03.65Bz, 67.90.+z, 05.30.Jp}

\begin{multicols}{2}

There has been a surge of interest in Bose-Einstein condensation
following the condensation of alkali gases in 1995 \cite{1995}.  A
defining feature of this recent work is the spatial inhomogeneity of
the gases due to their confinement in relatively small magnetic and
optical traps \cite{opticaltrap}.  Inhomogeneous Bose condensates also
occur in the presence of vortices
\cite{fetter72} and in the surface region of superfluid helium
\cite{he}.  Much theoretical work has focussed on modifications of
Bogoliubov's mean-field method to include the inhomogeneity arising,
in particular, from quadratic trapping potentials.  For particles with
a repulsive interaction the inhomogeneity leads to two important
effects. One is that the mean energy per particle depends on the
number of particles \cite{ground_state,bp}.  Wright {\it et al.}
\cite{wright} and Lewenstein {\it et al.} \cite{lewenstein,imamoglu}
have shown that this can lead to rapid phase diffusion and collapse of
the mean atomic field in a time much shorter than the lifetime of the
condensate and so mean-field descriptions of the condensate in these
cases remain valid only for relatively short times.  The other effect
is that the size of the ground-mode spatial wavefunction increases
with the number of the particles \cite{ground_state,bp} and so
changing the number of particles in a condensate without
correspondingly changing the spatial wavefunction will produce
occupation of excited trap levels.  This implies that a condensate
with a nonzero mean field, and thus with a spread in the number of
particles, is coupled in some way to the excited levels of the trap.
However, such particle-fluctuation-dependent couplings are not
included in descriptions of condensates to date.  The aim of this
Letter is to introduce a description of a condensed gas at zero
temperature that takes account of these two important effects.  The
new description depends only on the condensate possessing a large mean
atomic density without the necessity of a large mean field.  Since in
other respects it follows Bogoliubov's method it could be called a
{\it mean-density} Bogoliubov description \cite{compare}.  The
significant features that set it apart from conventional {\it
mean-field} Bogoliubov descriptions are that it remains valid even
when the mean field collapses and it includes an anomalous coupling
between the condensate and excited modes which depends on the
deviation of the condensate particle number from its mean value.  This
coupling could be used, in principle, to reduce the particle
fluctuations in the condensate.  The new description has important
consequences for studies of the quantum-field nature of condensates
such as state preparation \cite{state_preparation}, state
reconstruction
\cite{state_recon}, weakly coupled condensates \cite{weakly_coupled},
condensates as open systems \cite{open} and atom lasers
\cite{atom_lasers}.

We begin with the Hamiltonian for a weakly interacting boson gas in
second quantized form 
\begin{eqnarray}
\hat{\cal H} = \int d^3{\bf\text{r}}\hat\Psi^\dagger({\bf\text{r}})
   \left[{\cal L} +\frac{u}{2}\hat \Psi^\dagger({\bf\text{r}})
                \hat\Psi({\bf\text{r}})\right]\hat\Psi({\bf\text{r}})
\label{ham1}
\end{eqnarray}
where $\hat\Psi({\bf\text{r}})$ is the atomic field operator at
position ${\bf\text{r}}$, ${\cal L} \equiv -\frac{\hbar^2}{2m}
\nabla^2 + V({\bf\text{r}})$, $V({\bf\text{r}})$ is the trapping
potential, $u=4\pi\hbar^2 a_{\rm sc}/m$, $a_{\rm sc}$ is the $s$-wave
scattering length and $m$ is the atomic mass.  The objective is to
find an approximation to $\hat{\cal H}$ which is correct to order
unity.  It is convenient to expand $\hat\Psi({\bf\text{r}})$ in terms
of an orthonormal basis of spatial modes \cite{gardiner,walls}
\[
\hat\Psi({\bf\text{r}})=\sum^\infty_{n=0}\hat a_n \xi_n({\bf\text{r}})
\]
where $\hat a_n$ is the annihilation operator associated with the
spatial mode $\xi_n({\bf\text{r}})$ of the trapped gas with $[\hat
a_n,\hat a^\dagger_m]=\delta_{n,m}$, $[\hat a_n,\hat a_m]=0$ and $\int
d^3{\bf\text{r}}\xi_n^\ast({\bf\text{r}})\xi_m({\bf\text{r}})=\delta_{n,m}$.
The condensate is assumed to be in the ``ground'' mode represented by
$\hat a_n$ and $\xi_n$ with $n=0$ and so $\langle \hat a^\dagger_0\hat
a_0\rangle=N_0\gg 1$ whereas all ``excited'' modes are weakly
occupied, i.e. $\langle \hat a^\dagger_n \hat a_n\rangle\ll N_0$ for
$n\ge1$.  Also, the spread in the number of particles in the ground
mode is assumed to be no larger than Poissonian.  The effect of the
spatial-dependence on particle number is most clearly seen by writing
the atomic field operator in the form $\hat\Psi({\bf\text{r}})=\hat
a_0\xi_0({\bf\text{r}}) + \hat a_1\xi_1({\bf\text{r}}) + \hat
\psi_2({\bf\text{r}})$ where
\[
\hat \psi_n({\bf\text{r}})\equiv
          \sum_{m=n}^\infty \hat a_m \xi_m({\bf\text{r}})\ .
\]
Substituting into Eq. (\ref{ham1}) yields
\begin{equation}
\hat{\cal H}=\hat{\cal H}_{01}+\hat{\cal H}_{\rm rem}
\label{ham01rem}
\end{equation}
where $\hat{\cal H}_{01}$ contains all terms involving $\hat a_0$,
$\hat a_0^\dagger$, $\hat a_1$ and $\hat a_1^\dagger$ but not $\hat
\psi_2$ nor $\hat
\psi_2^\dagger$.

The most significant contribution to $\hat{\cal H}_{01}$ are the terms
involving only the operators $\hat a_0$ and $\hat a^\dagger_0$ and
these are given collectively by
\begin{eqnarray}
\hat H_g=\int d^3{\bf\text{r}}\hat a_0^\dagger\xi_0^\ast({\bf\text{r}})
  \left[{\cal L}+ \frac{u}{2}\hat a_0^\dagger\hat a_0|\xi_0({\bf\text{r}})|^2 
                            \right]\hat a_0\xi_0({\bf\text{r}})\ .
\label{Hg}
\end{eqnarray}
The normalised function $\xi_0({\bf\text{r}})$ that minimizes the mean
of this expression is a solution of the Gross-Pitaevskii equation
\begin{equation}
\left[{\cal L}
      + \overline{N} u|\xi_0({\bf\text{r}})|^2\right]
           \xi_0({\bf\text{r}})=\mu\xi_0({\bf\text{r}}) 
\label{gp}
\end{equation}
where $\overline{N}\equiv \langle\hat a_0^\dagger\hat a_0^\dagger\hat
a_0\hat a_0\rangle /\langle\hat a_0^\dagger\hat a_0\rangle\approx N_0$
and $\mu$ is the chemical potential.  Using Eq. (\ref{gp}) to remove
${\cal L}\xi_0({\bf\text{r}})$ from Eq. (\ref{Hg}) yields
\[
\hat{\cal H}_g=\mu\hat a_0^\dagger\hat a_0
    +\frac{u\alpha_2}{2}\hat a_0^\dagger(\hat a_0^\dagger\hat a_0
                               -2\overline{N})\hat a_0
\]
where $\alpha_n\equiv\int d^3{\bf\text{r}}
|\xi_0({\bf\text{r}})|^{2n}$.  This result shows that the
Gross-Pitaevskii equation (\ref{gp}) defines the minimum-energy
spatial wavefunction $\xi_0({\bf\text{r}})$ of the condensate for {\it
arbitrary states} of the condensate (for a given $\overline{N}$)
including those for which $\langle\hat a_0\rangle$, and thus the mean
field $\langle\hat\Psi\rangle\approx\langle\hat a_0\rangle\xi_0$, is
zero.

The next most significant contribution to $\hat{\cal H}_{01}$ is given
by terms which describe a coupling between the $n=0$ and $1$ modes.
For example consider the terms containing the product
$\xi_1^\ast({\bf\text{r}})\xi_0({\bf\text{r}})$ or its complex
conjugate to first order, i.e., the terms $\hat a_1^\dagger\hat
a_0\int d^3{\bf\text{r}}\xi_1^\ast({\bf\text{r}}){\cal
L}\xi_0({\bf\text{r}}) + \text{h.c.}$, which, on replacing ${\cal
L}\xi_0({\bf\text{r}})$ using Eq. (\ref{gp}), become
\begin{equation}
\hat a_1^\dagger\hat a_0\int d^3{\bf\text{r}}\xi_1^\ast({\bf\text{r}})\left[ 
       \mu-\overline{N} u|\xi_0({\bf\text{r}})|^2\right]
       \xi_0({\bf\text{r}})+\text{h.c.}
\label{t2}
\end{equation}
The contribution of the first term in the square brackets is zero
because $\xi_0({\bf\text{r}})$ and $\xi_1({\bf\text{r}})$ are
orthogonal.  If the contribution of the second term in square brackets
was also zero then there would be no coupling of this type between the
ground mode and the excited modes.  Therefore the part of the function
$|\xi_0({\bf\text{r}})|^2\xi_0({\bf\text{r}})$ which is orthogonal to
$\xi_0({\bf\text{r}})$ represents the spatial mode function coupled
{\it most strongly} to the ground mode.  $\xi_1({\bf\text{r}})$ can be
made to be this maximally-coupled mode by setting
\begin{equation}
\xi_1({\bf\text{r}})=
       \beta[|\xi_0({\bf\text{r}})|^2-\alpha_2]\xi_0({\bf\text{r}})
\label{xi1}
\end{equation}
where $\beta=(\alpha_3-\alpha^2_2)^{-1/2}$ is a normalisation
constant.  Fig. 1 illustrates the mode functions $\xi_0$ and
$\xi_1$ for typical experimental parameters.  With this choice for
$\xi_1({\bf\text{r}})$ expression (\ref{t2}) becomes
$-\frac{\overline{N}u}{\beta}\hat a_1^\dagger\hat a_0 + \text{h.c.}$.
Evaluating the integrals in the remaining terms in $\hat{\cal H}_{01}$
using $\xi_1({\bf\text{r}})$ defined by Eq. (\ref{xi1}) and the fact
that $\xi_0({\bf\text{r}})$ is a solution of Eq. (\ref{gp}) yields
\begin{figure}
\centerline{
\psfig{figure=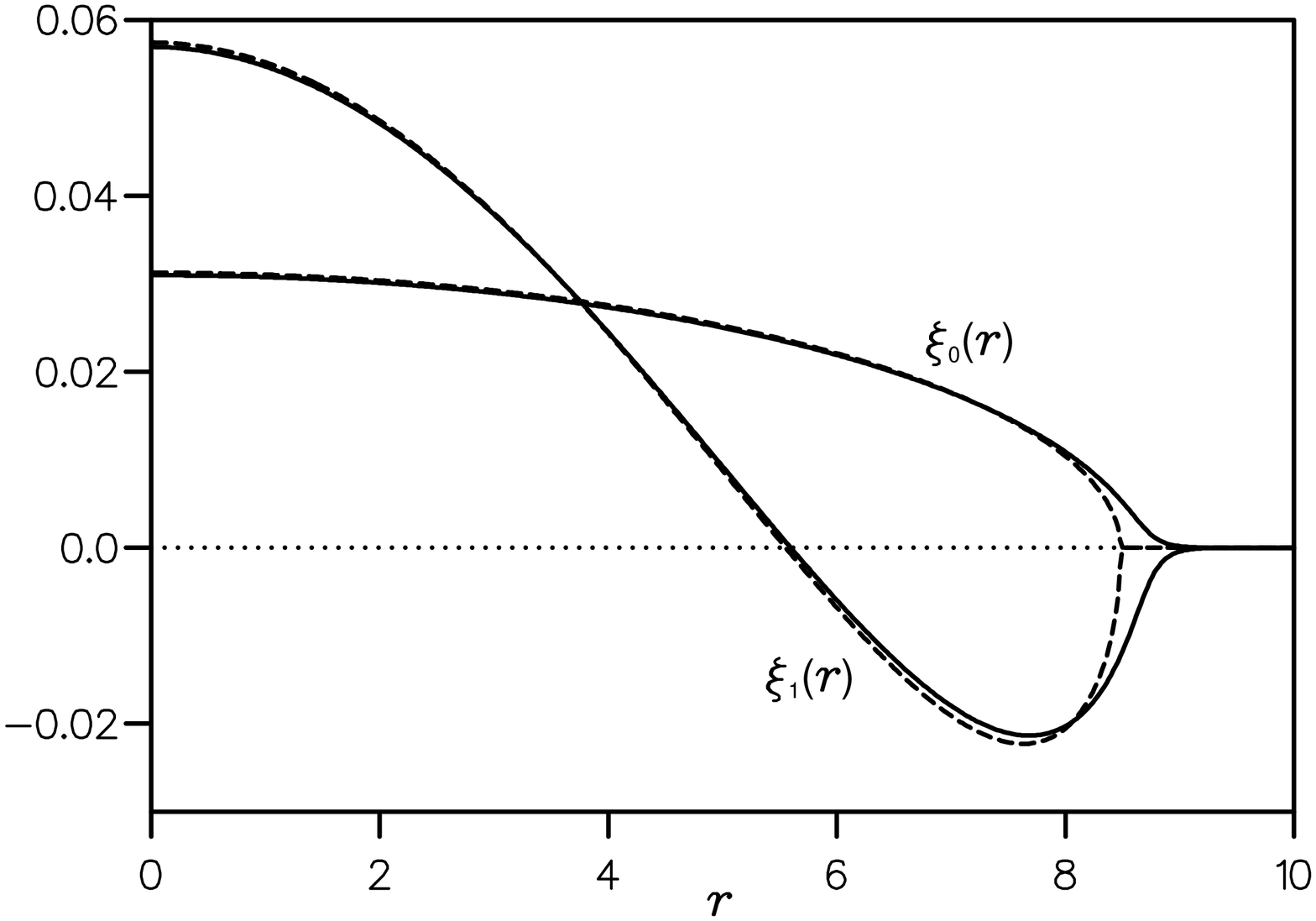,width=80mm,angle=-90,bbllx=19pt,bblly=40pt,bburx=500pt,bbury=750pt,clip=}
}
\noindent 
{\small FIG. 1. Spatial mode functions as a function of the radial coordinate
$r=|{\bf\text{r}}|$, in units where $r_0=\sqrt{\hbar/m\omega}=1$, for
a spherical harmonic trap of frequency 1000Hz with a condensate
containing $10^5$ rubidium atoms ($m=1.44\times 10^{-25}$ kg, $a_{\rm
sc}=10$ nm).  The solid and dashed curves are the Lambert function
\cite{ilinski} and Thomas-Fermi \cite{bp} approximations,
respectively.  }
\end{figure}
\begin{eqnarray}
\nonumber
\hat{\cal H}_{01}=\mu\hat a_0^\dagger\hat a_0
    +\frac{u\alpha_2}{2}\hat a_0^\dagger(\hat a_0^\dagger\hat a_0
                               -2\overline{N})\hat a_0\\
\nonumber
  +\left[\frac{u}{\beta}(\hat a_0^\dagger\hat a_0-\overline{N})
       \hat a_1^\dagger\hat a_0
    + \text{h.c.}\right] +\mu_1  \hat a_1^\dagger\hat a_1  \\
    +\gamma \hat a_1^\dagger\hat a_1(2\hat a_0^\dagger\hat a_0-\overline{N})
    +\frac{\gamma}{2}(\hat a_1^\dagger \hat a_1^\dagger\hat a_0\hat a_0 
    + \text{h.c.})
\label{H0}
\end{eqnarray}
where $\gamma=[\beta^2(\alpha_4-\alpha^3_2)-2\alpha_2]u$ and
$\mu_1=\mu+\frac{\hbar^2}{2m}\int d^3{\bf\text{r}}
\xi_1^\ast\beta[|\xi_0|^2\nabla^2\xi_0-\nabla^2(|\xi_0|^2\xi_0)]$. 
Terms of order one or lower in $\hat a_0$ or $\hat a_0^\dagger$ have been
neglected because their magnitude, being $O(1/\sqrt{N_0})$, is
relatively small.  The second term in Eq. (\ref{H0}) produces phase
diffusion of the mean field \cite{wright,walls}.  This term is quite
distinct from the corresponding ``momentum''-squared term in
Lewenstein and You's mean-field model \cite{lewenstein}.  In fact, the
``momentum''-squared term in \cite{lewenstein} is essentially a linearized
version of the above term for the special case where the condensate
has a large mean field and a phase of zero.  The next term in
Eq. (\ref{H0}) represents one of the main results of this work: it
describes {\it an anomalous coupling between the ground mode and mode
1 whose strength depends on the particle fluctuations in the ground
mode}.  This term embodies the coupling to the excited modes mentioned
in the opening paragraph.  The last three terms represent the
``usual'' terms for excited modes in the sense that replacing $\hat
a_0$ with $\sqrt{\overline{N}}$ gives the corresponding terms in the mean
field approach.

We can get some idea of the effect of the coupling between the ground
mode and mode 1 by considering the action of the Hamiltonian term
$\hat{\cal H}_{01}$ alone.  This ignores the effect of the other
excited modes, however, these effects should be small if the system is
in an approximate stationary state of $\hat{\cal H}_{\rm rem}$ in
Eq. (\ref{ham01rem}).  $\hat{\cal H}_{01}$ conserves the total
particle number and so one easy way to follow its effect is to
consider the evolution of a fixed number of particles, for example,
with initially $M$ particles in the ground mode and $0$ particles in
mode 1.  The value of $\overline{N}$ need not be equal to $M-1$,
i.e. $\xi_0({\bf\text{r}})$ need not be the lowest energy
wavefunction.  (The general solution for an arbitrary state
$|A\rangle$ with a spread in the number $M$ of particles in the ground
mode can then be obtained by an appropriate linear superposition of
the solutions for different values of $M$ but the same value of
$\overline{N}= \langle A|\hat a_0^\dagger\hat a_0^\dagger\hat a_0\hat
a_0|A\rangle /\langle A|\hat a_0^\dagger\hat a_0|A\rangle$.)  It is
not difficult to show for $M\gg 1$ that the evolution of the mean
particle number in mode 1 is given to a close approximation by
\begin{equation}
\langle \hat a_1^\dagger\hat a_1(t)\rangle
 = c_1\sin^2(\omega^\prime t)+c_2[\cos(\omega^\prime t)-1]^2
\label{mean_a1}
\end{equation}
which indicates that particles oscillate between the two modes.  Here
$c_1=[(M\gamma)^2+u^2(M-\overline{N})^2M/\beta^2]
/(\hbar\omega^\prime)^2$,
$c_2=u^2(M-\overline{N})^2M[(M-\overline{N})(\gamma-u\alpha_2)
+\mu_1-\mu]^2 /\beta^2(\hbar\omega^\prime)^4$ and
$(\hbar\omega^\prime)^2=[\gamma(2M-\overline{N})-(M-\overline{N})
u\alpha_2+\mu_1-\mu]^2-\gamma^2M^2$.  In the Thomas-Fermi regime
\cite{bp} the ground mode function for a harmonic trap of angular
frequency $\omega$ is approximated by
$\xi_0({\bf\text{r}})=[(\mu-|{\bf\text{r}}|^2
\hbar\omega/2r_0^2)/\overline{N}u]^{1/2}$
where $r_0\equiv\sqrt{\hbar/m\omega}$ for which $\mu=B\hbar\omega/2$,
$u\alpha_2=B\hbar\omega 2/7\overline{N}$, $u/\beta=B\hbar\omega
2/\sqrt{6}7\overline{N}$, $\gamma=B\hbar\omega 20/77\overline{N}$ and
$\mu_1=\mu+63\hbar\omega/2B$ where $B\equiv (15\overline{N}a_{\rm
sc}/r_0)^{2/5}$.  Taking $\gamma\approx u\alpha_2\approx 2u/\beta$ and
$\overline{N}\approx N_0$ leads to the simplifications
\begin{eqnarray*}
c_1 &\approx& [M+\frac{1}{4}(M-N_0)^2]
                     (a_{\rm sc}/r_0)^{4/5}(28 N_0^{1/5})^{-1}\\
c_2 &\approx& (M-N_0)^2(16M)^{-1}
\end{eqnarray*}
and $\omega^\prime\approx 4.0\sqrt{M/N_0}\omega\approx 4\omega$ for
$r_0/N_0 a_{\rm sc}\ll 1$.  For a condensate of rubidium atoms tightly
confined in a trap of frequency 1000Hz the number of atoms oscillating
between the two modes is of the order of $c_1\approx N_0^{4/5}/470$.
This can be a significant number for large condensates containing
$N_0\gg 500$ atoms, e.g. $N_0=10^6$ gives approximately
$200=\sqrt{N_0}/5$ atoms oscillating between the modes.  For a
Poisson-distributed occupation of the condensate a quarter of these
atoms are due to the anomalous term.

The dependence of the coefficients $c_1$ and $c_2$ on the difference
$M-N_0$ suggests a method for reducing particle fluctuations in the
ground mode that, despite its practical limitations at present, serves to
illustrate the nature of the anomalous coupling.  Imagine that after
evolving according to Eq. (\ref{mean_a1}) for a time
$\pi/\omega^\prime$ the population of mode 1 is rapidly depleted by
some means.  Although difficult in practice, in principle it
would be possible to do this without simultaneously depleting the
ground mode using the method proposed by Walsworth and You
\cite{walsworth}.  This removes $4c_2$ particles, on average, from the
system. After repeating this process many
times the population of the ground mode either approaches
$N_0$ or decays rapidly to zero
depending on whether initially $M>N_0$ or $M<N_0$,
respectively.  Evidently this cyclic process is a measurement of the
occupation of the ground mode that distinguishes only between initial
occupations $M$ that are above and below $N_0$.  Thus, by
linearity, if the ground mode begins in a pure state which has, e.g.,
a symmetrical spread in particle number $M$ about the mean
$N_0$ it will eventually be transformed into a
state which has, with equal probability, either approximately
$N_0$ or zero particles.  This method, therefore,
has a $50\%$ efficiency in reducing particle fluctuations in the
ground mode.

The final task in developing this mean-density Bogoliubov description
is to identify the quasiparticle modes.  The expansion of the
Hamiltonian in Eq. (\ref{ham01rem}) was chosen to express most clearly
the coupling between the ground mode and mode 1.  It is more
appropriate here, however, to write the atomic field operator as
$\hat\Psi({\bf\text{r}})=\hat a_0\xi_0({\bf\text{r}}) +
\hat\psi_1({\bf\text{r}})$ and expand the Hamiltonian to second order
in $\hat\psi_1({\bf\text{r}})$.  Transforming off the uninteresting
chemical potential gives the Hamiltonian in grand canonical form,
$\hat{\cal H}^\prime=\hat{\cal H}-\int
d^3{\bf\text{r}}\mu\hat\Psi^\dagger({\bf\text{r}})\hat\Psi({\bf\text{r}})$,
for which the required expansion can be written as
\begin{equation}
\hat{\cal H}^\prime=\hat{\cal H}^\prime_g+\hat{\cal H}_{ge}+\hat{\cal H}_e
\label{Hquasiparticle}
\end{equation}
where $\hat{\cal H}^\prime_g$, $\hat{\cal H}_{ge}$ and $\hat{\cal
H}_e$ are, respectively, independent, linear and bilinear in
$\hat\psi_1({\bf\text{r}})$ and $\hat\psi_1^\dagger({\bf\text{r}})$.
Choosing $\xi_0({\bf\text{r}})$ to minimize the mean of $\hat{\cal
H}_g$, as before, yields
\begin{equation}
\hat{\cal H}^\prime_g=
        \frac{u\alpha_2}{2}\hat a_0^\dagger(\hat a_0^\dagger\hat a_0
                               -2\overline{N})\hat a_0 \ .
\label{Hg_}
\end{equation}
The foregoing analysis also shows that $\hat{\cal H}_{ge}$ represents
the coupling between the ground mode and mode 1 as given by the third
term on the right side of Eq. (\ref{H0}), i.e.
\begin{equation}
\hat{\cal H}_{ge}=\frac{u}{\beta}
(\hat a_0^\dagger\hat a_0-\overline{N})\hat a_1^\dagger\hat a_0+ \text{h.c.}\ .
\label{Hge}
\end{equation}
It is useful to consider this coupling term as a perturbation on a
system whose Hamiltonian is given by $\hat{\cal H}^\prime_g+\hat{\cal
H}_e$.

In this perturbative approach the quasiparticle modes are those that
diagonalize $\hat{\cal H}_e$,
\begin{eqnarray}
\nonumber
\hat{\cal H}_e=
  \int d^3{\bf\text{r}}\left[\hat\psi^\dagger_1({\cal L}-\mu)\hat\psi_1 + 
   (\frac{u}{2}\hat\psi^\dagger_1\hat\psi^\dagger_1\hat a_0\hat a_0\xi_0^2
    +\text{h.c.})\right.\\
  +\left.2u\hat\psi^\dagger_1\hat\psi_1
            \hat a_0^\dagger\hat a_0|\xi_0|^2\right]\ .
\label{He}
\end{eqnarray}
The diagonalization can be carried out by generalizing the method used
by Fetter \cite{fetter72} to accommodate the ground mode operators
$\hat a_0^\dagger$, $\hat a_0$ in Eq. (\ref{He}) which are absent from
Fetter's mean-field treatment.  Since $\hat{\cal H}_e$ conserves total
particle number the required Bogoliubov transformation must also
conserve particle number.  We therefore adopt the ansatz
\cite{gardiner_castin}
\begin{equation}
\frac{\hat a^\dagger_0}{\sqrt{\overline{N}}}\hat\psi_1({\bf\text{r}})
      =\sum_{k=1}^\infty
  [U_k({\bf\text{r}})\hat b_k-V^\ast_k({\bf\text{r}})\hat b^\dagger_k]
\label{defineb}
\end{equation}
where $\{\hat b_k\}_k$ are a set of independent quasiparticle
operators with $[\hat b_k,\hat b_j]=0$ and $[\hat b_k,\hat
b_j^\dagger]=\delta_{k,j}$.  Eq. (\ref{defineb}) implies that
$U_k({\bf\text{r}})$ and $V_k({\bf\text{r}})$ are orthogonal to
$\xi_0({\bf\text{r}})$; this fact can be made more transparent by
defining
\begin{mathletters}
\label{uv}
\begin{eqnarray}
U_k({\bf\text{r}})
  =u_k({\bf\text{r}})-\xi_0({\bf\text{r}})
     \int d^3{\bf\text{r}} \xi_0^\ast({\bf\text{r}})u_k({\bf\text{r}}) \\
V_k({\bf\text{r}})
  =v_k({\bf\text{r}})-\xi_0({\bf\text{r}})
     \int d^3{\bf\text{r}} \xi_0^\ast({\bf\text{r}})v_k({\bf\text{r}})
\end{eqnarray}
\end{mathletters}
and working instead with $u_k({\bf\text{r}})$ and $v_k({\bf\text{r}})$
which are elements of the whole space spanned by
$\{\xi_n({\bf\text{r}})\}_{n=0,1,\ldots}$.  Since the magnitude of
$\hat{\cal H}_e$ is of order unity one can make approximations of
relative error of order $1/\sqrt{N_0}$ such as replacing the operator
$\hat a^\dagger_0\hat a_0$ with its mean value $N_0$.  Multiplying the
first term in Eq. (\ref{He}) by $\hat a^\dagger_0\hat
a_0/\overline{N}=1+O(1/\sqrt{N_0})$, which is effectively unity here,
and then substituting for $\frac{\hat
a_0^\dagger}{\sqrt{\overline{N}}}\hat\psi_1({\bf\text{r}})$ and its
hermitian conjugate using Eq. (\ref{defineb}) yields an approximate
expression for $\hat{\cal H}_e$ which does not involve $\hat a_0$ nor
$\hat a_0^\dagger$.  It can be shown using an analysis similar to
that of Fetter
\cite{fetter72} that
\begin{equation}
\hat{\cal H}_e=C+\sum_{k=1}^\infty\hbar\omega_k\hat b^\dagger_k\hat b_k 
\label{He2}
\end{equation}
to order unity, where $\omega_k$, $u_k({\bf\text{r}})$ and
$v_k({\bf\text{r}})$ satisfy the Bogoliubov-de Gennes equations
\begin{eqnarray*}
\tilde{{\cal L}}u_k({\bf\text{r}})
      -u\overline{N}|\xi_0({\bf\text{r}})|^2 v_k({\bf\text{r}})=
                   \hbar\omega_k u_k({\bf\text{r}})\\
\tilde{{\cal L}}v_k({\bf\text{r}})
      -u\overline{N}|\xi_0({\bf\text{r}})|^2 u_k({\bf\text{r}})=
                   -\hbar\omega_k v_k({\bf\text{r}})
\end{eqnarray*}
with $\tilde{{\cal L}}= -\frac{\hbar^2}{2m} \nabla^2 +
V({\bf\text{r}})-\mu+2u\overline{N}|\xi_0({\bf\text{r}})|^2$, and
where $C=-\sum_{k,j=1}^\infty\hbar\omega_k\int d^3{\bf\text{r}}
[V_j({\bf\text{r}})v_k^\ast({\bf\text{r}})+ {\rm c.c.}]/2$.

It is now possible to find the decomposition of mode 1 in terms of the
quasiparticle modes via the overlap
\[
\frac{\hat a^\dagger_0}{\sqrt{\overline{N}}}\hat a_1=\int d^3{\bf\text{r}} 
    \xi_1^\ast({\bf\text{r}})
    \frac{\hat a^\dagger_0}{\sqrt{\overline{N}}}\hat\psi_1({\bf\text{r}})\ .
\]
Using Eqs. (\ref{defineb}) and (\ref{uv}) with the approximate
expressions for $u_k$ and $v_k$ in the Thomas-Fermi regime found by
\"Ohberg {\it et al.} \cite{ohberg} for the physical parameters
used in Fig. 1 gives $\frac{\hat a_0^\dagger}{\sqrt{\overline{N}}}\hat
a_1\approx 1.755\hat b_1-1.443\hat b_1^\dagger - 0.986(\hat b_2-\hat
b_2^\dagger)$, that is, mode 1 is composed of the first two
quasiparticle modes only \cite{note_a1}.

In conclusion, this Letter introduces a new description of
inhomogeneous Bose-condensed gases which is based on Bogoliubov's
method but assumes a large {\it mean density}
$\langle\hat\Psi^\dagger\hat\Psi\rangle\approx\langle\hat
a_0^\dagger\hat a_0\rangle|\xi_0|^2$ instead of a large mean field.
The description remains valid, therefore, when the mean field
collapses due to phase diffusion.  Eq. (\ref{Hquasiparticle}) together
with Eqs. (\ref{Hg_}), (\ref{Hge}) and (\ref{He2}) represent its main
result: an approximate Hamiltonian which is valid for arbitrary states
of the condensate and which includes an anomalous coupling between the
ground and quasiparticle modes.  This coupling, which has not been
described previously, is due to the dependence of the ground-mode
spatial wavefunction on particle number. It vanishes on taking the
mean of $\hat{\cal H}^\prime$ with respect to a coherent state of the
ground mode, i.e. it vanishes in the mean-field approach.  The
mean-density description thus accommodates the two important effects
of inhomogeneous condensed gases that have come to light recently:
rapid phase diffusion and the dependence of the ground-mode spatial
wavefunction on particle number.

It is a pleasure to acknowledge helpful discussions with A. Durrant,
S. Hopkins, U. Leonhardt, V. Penna, D. Richards and T.B. Smith
concerning this work.


\end{multicols} 

\begin{references}
\bibitem{1995}
M.H. Anderson {\it et al.}, Science {\bf 269}, 198 (1995);
C.C. Bradley {\it et al.}, Phys. Rev. Lett. {\bf 75}, 1687 (1995);
K.B. Davis {\it et al.}, Phys. Rev. Lett. {\bf 75}, 3969 (1995).

\bibitem{opticaltrap}
D.M. Stamper-Kurn {\it et al.}, Phys. Rev. Lett. {\bf 80}, 2027 (1998).

\bibitem{fetter72}
A.L. Fetter, Ann. Phys. (N.Y.) {\bf 70}, 67 (1972).

\bibitem{he}
A. Griffin {\it et al.}, Phys. Rev. Lett. {\bf 76}, 259 (1996).

\bibitem{ground_state}
M. Edwards {\it et al.}, Phys. Rev. A {\bf 53}, R1950 (1996);
F. Dalfovo and S. Stringari, Phys. Rev. A {\bf 53}, 2477 (1996).

\bibitem{bp}
G. Baym and C.J. Pethick, Phys. Rev. Lett. {\bf 76}, 6 (1996).

\bibitem{wright}
E.M. Wright {\it et al.}, Phys. Rev. Lett. {\bf 77}, 2158 (1996).

\bibitem{lewenstein}
M. Lewenstein {\it et al.},  Phys. Rev. Lett. {\bf 77}, 3489 (1996).

\bibitem{imamoglu}
A. Imamo$\bar{\text{g}}$lu {\it et al.}, Phys. Rev. Lett. {\bf 78},
2511 (1997).

\bibitem{compare}
It is important to distinguish the description introduced here from
that of Gardiner \cite{gardiner}, and equivalently Castin and Dum
\cite{castin}, which could also be called a mean-density description.  
The main difference is that the latter does not
allow for a coherent spread in condensate particle number (and thus 
a nonzero mean field) which underpins the main result of this work.

\bibitem{state_preparation}
R. Graham {\it et al.}, Phys. Rev. A {\bf 57}, 493 (1998); J.I. Cirac
{\it et al.}, Phys. Rev. A {\bf 57}, 1208 (1998).

\bibitem{state_recon}
E.L. Bolda {\it et al.}, Phys. Rev. Lett. {\bf 79}, 4719 (1997);
R. Walser, {\it ibid.}, 4724 (1997).

\bibitem{weakly_coupled}
M.J. Steel {\it et al.}, Phys. Rev. A {\bf 57}, 2920 (1998).

\bibitem{open}
J. Anglin, Phys. Rev. Lett. {\bf 79}, 6 (1997).

\bibitem{atom_lasers}
See e.g. H.M. Wiseman, Phys. Rev. A {\bf 56}, 2068 (1997).


\bibitem{gardiner}
C.W. Gardiner,  Phys. Rev. A {\bf 56}, 1414 (1997).

\bibitem{walls}
E.M. Wright {\it et al.}, Phys. Rev. A {\bf 56}, 591 (1997).

\bibitem{walsworth}
R. Walsworth and L. You, Phys. Rev. A {\bf 56}, 555 (1997).

\bibitem{gardiner_castin}
This definition of the quasiparticle operators is equivalent to that
of Refs. \cite{gardiner,castin}; see also note \cite{compare}.
 
\bibitem{ohberg}
P. \"Ohberg {\it et al.}, Phys. Rev. A {\bf 56}, R3346 (1997).

\bibitem{note_a1}
Note that $[\hat a_0^\dagger\hat a_1,\hat a_0\hat
a_1^\dagger]/\overline{N}=(\hat a_0^\dagger\hat a_0-\hat a_1\hat
a_1^\dagger)/\overline{N}= 1+ O(1/\sqrt{N_0})$ and so effectively
$\hat a_1\sim 1.755\hat b_1-1.443\hat b_1^\dagger - 0.986(\hat
b_2-\hat b_2^\dagger)$.

\bibitem{castin}
Y. Castin and R. Dum, Phys. Rev. A {\bf 57}, 3008 (1998)

\bibitem{ilinski}
N. Ilinski and A. Moroz, J. Res. Natl. Inst. of Standards and
Technology {\bf 101}, 567 (1996).

\end{references}
\end{document}